\documentclass[letter,twocolumn]{jpsj3}
\usepackage{txfonts}
\usepackage{graphicx}
\usepackage{dcolumn}
\usepackage{color}
\usepackage{array,multirow,makecell}
\usepackage{bm}
\usepackage{nicefrac}
\usepackage{multicol,blindtext}

\title{Magnetic-Field-Induced Phenomena in the Paramagnetic Superconductor UTe$_{2}$}

\author{William \textsc{Knafo}$^{1}$, Michal \textsc{Vali\v{s}ka}$^{2}$, Daniel \textsc{Braithwaite}$^{2}$, G\'{e}rard \textsc{Lapertot}$^{2}$, Georg \textsc{Knebel}$^{2}$, Alexandre \textsc{Pourret}$^{2}$, Jean-Pascal \textsc{Brison}$^{2}$, Jacques \textsc{Flouquet}$^{2}$, Dai \textsc{Aoki}$^{2,3}$}

\inst{
$^{1}$ Laboratoire National des Champs Magn\'{e}tiques Intenses, UPR 3228, CNRS-UPS-INSA-UGA, 143 Avenue de Rangueil, 31400 Toulouse, France\\
$^{2}$ Universit\'{e} Grenoble Alpes and CEA, IRIG-PHELIQS, F-38000 Grenoble, France\\
$^{3}$ Institute for Materials Research, Tohoku University, Ibaraki 311-1313, Japan\\
}

\abst{
We present magnetoresistivity measurements on the heavy-fermion superconductor UTe$_{2}$ in pulsed magnetic fields $\mu_0H$ up to 68~T and temperatures $T$ from 1.4 to 80~K. Magnetic fields applied along the three crystallographic directions $\mathbf{a}$ (easy magnetic axis), $\mathbf{b}$, and $\mathbf{c}$ (hard magnetic axes), are found to induce different phenomena - depending on the field direction - beyond the low-field suppression of the superconducting state. For $\mathbf{H}\parallel\mathbf{a}$, a broad anomaly in the resistivity is observed at $\mu_0H^*\simeq10$~T and $T = 1.4$~K. For $\mathbf{H}\parallel\mathbf{c}$, no magnetic transition nor crossover are observed. For $\mathbf{H}\parallel\mathbf{b}$, a sharp first-order-like step in the resistivity indicates a metamagnetic transition at the field $\mu_0H_m \simeq 35$~T. When the temperature is raised signature of first-order metamagnetism is observed up to a critical endpoint at $T_{CEP}\simeq7$~K. At higher temperatures a crossover persists up to 28~K, i.e., below the temperature $T_\chi^{max} = 35$~K where the magnetic susceptibility is maximal. A sharp maximum in the Fermi-liquid quadratic coefficient $A$ of the low-temperature resistivity is found at $H_m$. It indicates an enhanced effective mass associated with critical magnetic fluctuations, possibly coupled with a Fermi surface instability. Similarly to the URhGe case, we show that UTe$_{2}$ is a candidate for field-induced reentrant superconductivity in the proximity of $H_m$.
}

\begin{document}
\maketitle

The microscopic coexistence of unconventional superconductivity and ferromagnetism in the U-based compounds (UGe$_2$,\cite{Saxena2000} URhGe,\cite{Aoki2001} UCoGe,\cite{Huy2007} and UIr \cite{Akazawa2004}) is one of the most exciting subjects of research in strongly correlated electron systems.\cite{Aoki2012,Aoki2019} Aside from this surprising co-existence, one of the main points of interest comes from the mechanism of superconductivity that is suspected to stem from ferromagnetic fluctuations. One remarkable consequence of this is that the application of an external field can significantly modify the strength of the pairing mechanism, depressing it or on the contrary enhancing it, depending on the field orientation in relation to the magnetic anisotropy of the system.\cite{Wu2017} In the case of URhGe under a magnetic field $\mathbf{H}$ applied along the intermediate magnetic axis $\mathbf{b}$, this phenomenon shows up spectacularly as re-entrant superconductivity at the metamagnetic transition $\mu_0H_m =12$~T.\cite{Levy2005} The metamagnetic transition is governed by a collapse of ferromagnetic order (with magnetic moments along $\mathbf{c}$ driven by a rotation of the magnetic moments to the $\mathbf{b}$-axis, as shown by a jump in the $\mathbf{b}$-axis magnetization \cite{Levy2005,Hardy2011}. A modification of the Fermi surface at $H_m$ was evidenced by Shubnikov-de-Haas and thermoelectric power experiments \cite{Yelland2011,Gourgout2016}. A maximum in the quadratic temperature-dependence of the normal-state resistivity also indicates an increase of the effective mass, and thus, of the magnetic fluctuations \cite{Gourgout2016,Miyake2008,Miyake2009}. These enhanced critical magnetic fluctuations are suspected to drive the field-induced superconducting pairing at $H_m$. In the case of UCoGe in a field applied along its intermediate magnetic axis $\mathbf{b}$, an S-shape at $\simeq 15$~T in the temperature dependence of the superconducting critical field $H_{c2}$ was also identified as a signature of field-induced superconductivity \cite{Aoki2009}. However, reentrance of superconductivity was found to be disconnected from the metamagnetic transition observed at $\mu_0H_m\simeq 50$~T in this system \cite{Knafo2012}.

Superconductivity was recently discovered in the heavy-fermion paramagnet UTe$_{2}$ at temperatures below $T_{sc}=1.6$~K.\cite{Ran2019} UTe$_{2}$ has an orthorhombic crystal structure with the Immm space group and its room-temperature lattice parameters are $a = 4.1611(7)$~{\AA},  $b  =  6.1222(9)$~{\AA} and  $c = 13.955(2)$~{\AA}.\cite{Ikeda2006} The shortest inter-uranium distance $d_{UU} = 3.7801$~{\AA} is along the $\mathbf{c}$-direction, forcing the magnetic easy axis to be perpendicular (here along the $\mathbf{a}$-axis\cite{Ikeda2006,Ran2019}) as in the vast majority of U-based intermetallics.\cite{Sechovsky1998} A strongly-anisotropic superconducting upper critical field $H_{c2}$ exceeding the Pauli limit for the three main crystallographic directions indicates spin-triplet superconductivity. For a magnetic field applied along $\mathbf{b}$, the shape of the $H_{c2}$ phase boundary is strongly sample-dependent,\cite{Aoki2019b} exhibiting  an upturn in some of the highest-quality samples. For all samples, an anomalous shape of $H_{c2}$ can be described assuming a field-induced enhancement of the pairing strength.\cite{Aoki2019b} Although the field-behavior of superconductivity in UTe$_{2}$ is quite similar to that of URhGe and UCoGe, \cite{Levy2005,Aoki2009} there is a major difference between these systems; UTe$_{2}$ is paramagnetic while URhGe and UCoGe are ferromagnetic. However, UTe$_{2}$ seems to be on the verge of ferromagnetism, as indicated by the low-temperature enhancement (at $T>T_{sc}$) of its magnetic susceptibility for $\mathbf{H} \parallel \mathbf{a}$ \cite{Ran2019}. Similarly to the URhGe and UCoGe cases, the magnetic susceptibility of UTe$_{2}$ shows a pronounced anisotropy at low temperature. For $\mathbf{H} \parallel \mathbf{b}$, which is the intermediate magnetic axis for the three compounds at high temperature ($\textbf{b}$ becomes the hardest magnetic axis at low temperature in UTe$_2$), the fact that a maximum in the magnetic susceptibility is observed at $T_\chi^{max} = 35$~K in UTe$_{2}$ \cite{Ran2019} indicates that metamagnetism is expected to occur, as well as it was observed in URhGe and UCoGe \cite{Knafo2012}.

In this work, we have performed magnetoresistivity measurements on UTe$_{2}$ single crystals in pulsed magnetic fields up to 68~T applied along the three crystallographic directions $\mathbf{a}$, $\mathbf{b}$ and $\mathbf{c}$, at temperatures from 1.4 to 80~K. For $\mathbf{H} \parallel \mathbf{b}$, we find evidence for a metamagnetic transition accompanied by a sharp jump of the resistivity and an enhancement of the effective mass at the field $\mu_0H_m = 35$~T. A broad crossover is also observed at a field $\mu_0H^*\simeq10$~T applied along $\mathbf{a}$, while no signature of magnetic transition or crossover is found for $\mathbf{H} \parallel \mathbf{c}$. In the light of previous studies made on the URhGe and UCoGe, a relation between metamagnetism and a possible field-induced enhancement of superconductivity is proposed.

Single crystals of UTe$_{2}$ were prepared by the chemical vapor transport method with similar parameters as described in Ref. \cite{Ran2019}. Their structure and orientation was checked by single-crystal X-ray diffraction. A sharp bulk transition at $T_{sc}=1.5$~K was indicated from specific heat measurements, while zero-resistivity at temperatures below $T_{sc}$ was confirmed by zero-field AC resistivity measurements. Magnetoresistance measurements were performed at the Laboratoire National des Champs Magn\'{e}tiques Intenses (LNCMI) in Toulouse under long-duration (30~ms raise and 100~ms fall) pulsed magnetic fields up to 68~T. Standard four-probe method with currents $\mathbf{I}\parallel\mathbf{a}$ at a frequency of 20-70~kHz and digital lock-in detection was used. Three samples of residual resistivity ratios $\rho_{xx}(T=300\textrm{~K})/ \rho_{xx}(T=2\textrm{~K})\simeq25$ were measured with three orientations of the magnetic field $\mathbf{H} \parallel \mathbf{a}$,  $\mathbf{b}$, and  $\mathbf{c}$.

\begin{figure}[t]
\includegraphics[width=.95\columnwidth]{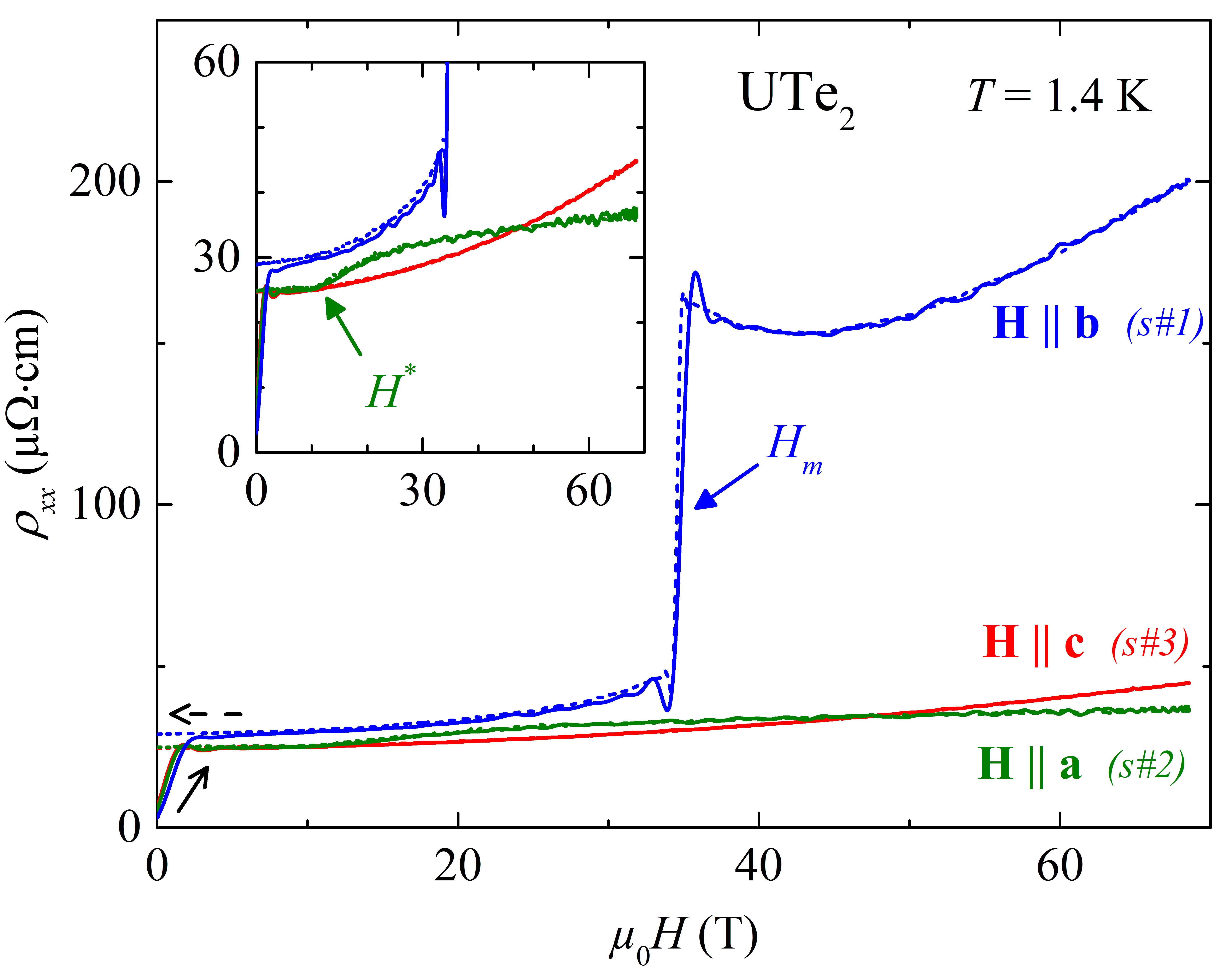}
\caption{(Color online) Magnetoresistivity versus field of UTe$_{2}$ at $T = 1.4$~K for fields $\mu_0\mathbf{H} \parallel \mathbf{a}$,  $\mathbf{b}$, and $\mathbf{c}$ up to 68~T. Both field-up (solid lines) and field-down (dotted line) sweeps are plotted. Black arrows indicate the time direction during the field pulses. Inset focuses on the small resistivity variations observed for $\mathbf{H} \parallel \mathbf{a},\mathbf{c}$.}
\label{Fig1}
\end{figure}

\begin{figure}[t]
\includegraphics[width=1.02\columnwidth]{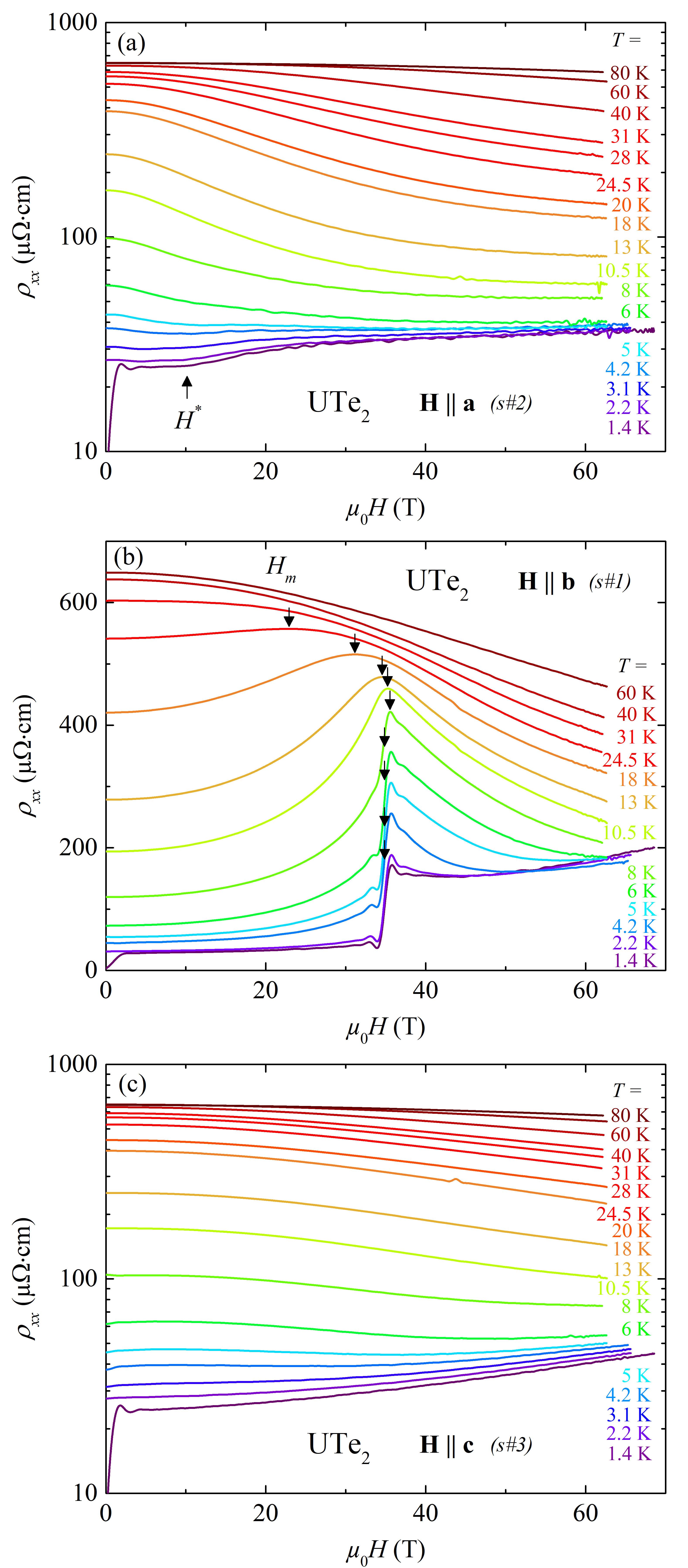}
\caption{(Color online) Magnetoresistivity versus field of UTe$_{2}$ (a) for $\mu_0\mathbf{H} \parallel \mathbf{a}$ (semi-logarithmic scale), (b) for $\mu_0\mathbf{H} \parallel \mathbf{b}$, and (c) for $\mu_0\mathbf{H} \parallel \mathbf{c}$ (semi-logarithmic scale) up to 68~T at temperatures from 1.4 to 80~K. Only field-up sweeps are shown.}
\label{Fig2}
\end{figure}

Figure \ref{Fig1} shows the magnetoresistivity $\rho_{xx}(H)$ of UTe$_{2}$ at $T = 1.4$~K, i.e., just below $T_{sc}$, for magnetic fields $\mathbf{H}$ applied along $\mathbf{a}$,  $\mathbf{b}$, and  $\mathbf{c}$. Both up and down field-sweeps are shown. All field-up curves start in the superconducting and zero-resistance state while the field-down curves end with non-zero resistivity, pointing to a small sample heating during the field-pulse. For $\mathbf{H}\parallel\mathbf{a}$, we detect a broad anomaly associated with a small increase in $\rho_{xx}$, whose inset can be defined at the field $\mu_0H^*\simeq10$~T. For $\mathbf{H} \parallel \mathbf{b}$, a sharp step-like variation of the resistivity, which increases by about a factor 4, is observed  at $\mu_0H_m = 35$~T. From our resistivity measurements there is little doubt that the feature at $H_m$ corresponds to a first-order metamagnetic transition. This has been confirmed by high-field magnetization measurements.\cite{Miyake2019} For rising fields, additional peaks before and after the large step in $\rho_{xx}(H)$ at $H_m$ result from experimental artefacts (as discussed in the Supplementary Materials \cite{SM}) and will not be discussed further. Contrary to the cases with a field $\mathbf{H} \parallel \mathbf{a}$ and $\mathbf{b}$, no anomaly is found in a field $\mu_0\mathbf{H} \parallel \mathbf{c}$ up to 68~T and the low-temperature resistivity simply follows a orbital $H^2$ variation controlled by the field-induced cyclotron motion of carriers in a compensated metal. A $H^2$ behavior is also present in the low-temperature high-field resistivity for $\mathbf{H} \parallel \mathbf{a}$ and $\mathbf{b}$ (see Supplementary Materials \cite{SM}).

Figure \ref{Fig2} shows the magnetoresistivity $\rho_{xx}(H)$ of UTe$_{2}$ for a large set of temperatures from 1.4 to 80~K for the three orientations of field. For $\mathbf{H} \parallel \mathbf{a}$, the crossover $H^*$ shifts to lower fields with increasing temperature and we rapidly lose its trace above 4.2~K.  While $\rho_{xx}(H)$ increases monotonously at $T = 1.4$~K, it decreases monotonously at all  temperatures $T \geq 6$~K. For $\mathbf{H} \parallel \mathbf{b}$, the step-like anomaly at $H_m$ keeps its sharp character and position in temperatures up to 6~K. At temperatures $T\geq8$~K, a crossover has replaced the step and is characterized by fast broadening and shift down to lower fields at higher temperatures, where it can be traced up to 28~K. $H_m$ is associated with a sharp maximum in the field-derivative $\partial\rho_{xx}/\partial H$ of the resistivity for $T \leq 6$~K and with a broad maximum of $\rho_{xx}(H)$ for $T \geq 8$~ K (see Supplementary Materials \cite{SM}). For $T \leq 6$~K, the first-order character of the transition is accompanied by an hysteresis, whose maximal width reaches $\Delta(\mu_0H)=0.3$~T at low-temperature. By warming up, the first-order transition ends at the critical endpoint characterized by the temperature $T_{CEP}=7$~K. This observation has been confirmed by magnetization measurements \cite{Miyake2019}. At temperatures above 30~K, a continuous decrease of $\rho_{xx}(H)$ is observed. Finally, for $\mathbf{H} \parallel \mathbf{c}$, the shape of the temperature-dependence of $\rho_{xx}(H)$, with a low-temperature monotonous increase and a high-temperature monotonous decrease, looks rather similar to that observed for $\mathbf{H} \parallel \mathbf{a}$. However, no trace of magnetic transition or crossover is observed for $\mathbf{H} \parallel \mathbf{c}$.

\begin{figure}[t]
\includegraphics[width=1\columnwidth]{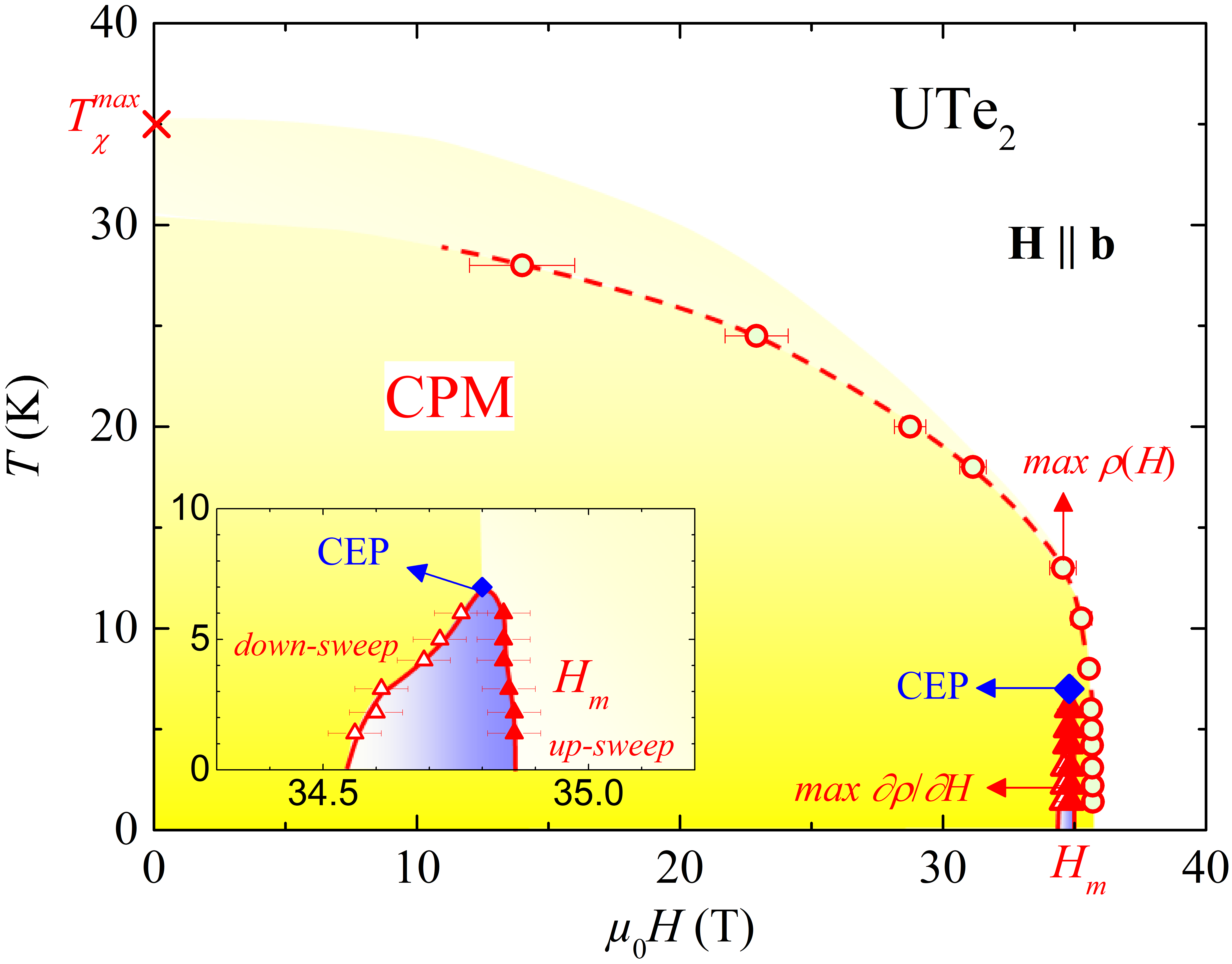}
\caption{(Color online) Magnetic-field temperature phase diagram of UTe$_{2}$ for $\mathbf{H} \parallel \mathbf{b}$. Inset focuses on the low-temperature hysteresis at $H_m$ ending at the critical endpoint at $T_{CEP}\simeq7$~K. The superconducting phase is not represented here.}
\label{Fig3}
\end{figure}

Figure \ref{Fig3} shows the resulting magnetic-field-temperature phase diagram of UTe$_{2}$ for $\mathbf{H} \parallel \mathbf{b}$. A striking feature is that the value of $H_m$ is almost temperature-independent as long as $T \leq T_{CEP}=7$~K, i.e., as long as the metamagnetic transition induces a sharp first-order-like step in $\rho_{xx}(H)$. When the temperature is increased above $T_{CEP}$, $H_m$ falls down and is characterized by a broad maximum in $\rho_{xx}(H)$. The trace of $H_m$ is lost at temperatures $T \geq 30$~K, which roughly coincides with the crossover temperature $T_\chi^{max} = 35$~K, where a broad maximum in the magnetic susceptibility marks the onset of a correlated-paramagnetic (CPM) regime. In many heavy-fermion paramagnets, and in a few antiferromagnets for $T > T_N$ and ferromagnets for $T > T_C$, a CPM regime is delimited by the borderlines $T_\chi^{max}$ (in the limit of zero-magnetic-field) and $H_m$ (in the limit of zero-temperature) \cite{Aoki2013,Aoki2012b,Knafo2017,Pospisil2017,Pospisil2018}. Remarkably, a simple relation between $T_\chi^{max}$  and $H_m$ (1~K~$\leftrightarrow$~1~T) holds for most of these systems.\cite{Aoki2013} This correspondence suggests a common magnetic energy scale controlling $T_\chi^{max}$ and $H_m$, and thus, the electronic correlations leading to strong quantum magnetic fluctuations and to a high effective mass in the CPM regime.

\begin{figure}[t]
\includegraphics[width=1\columnwidth]{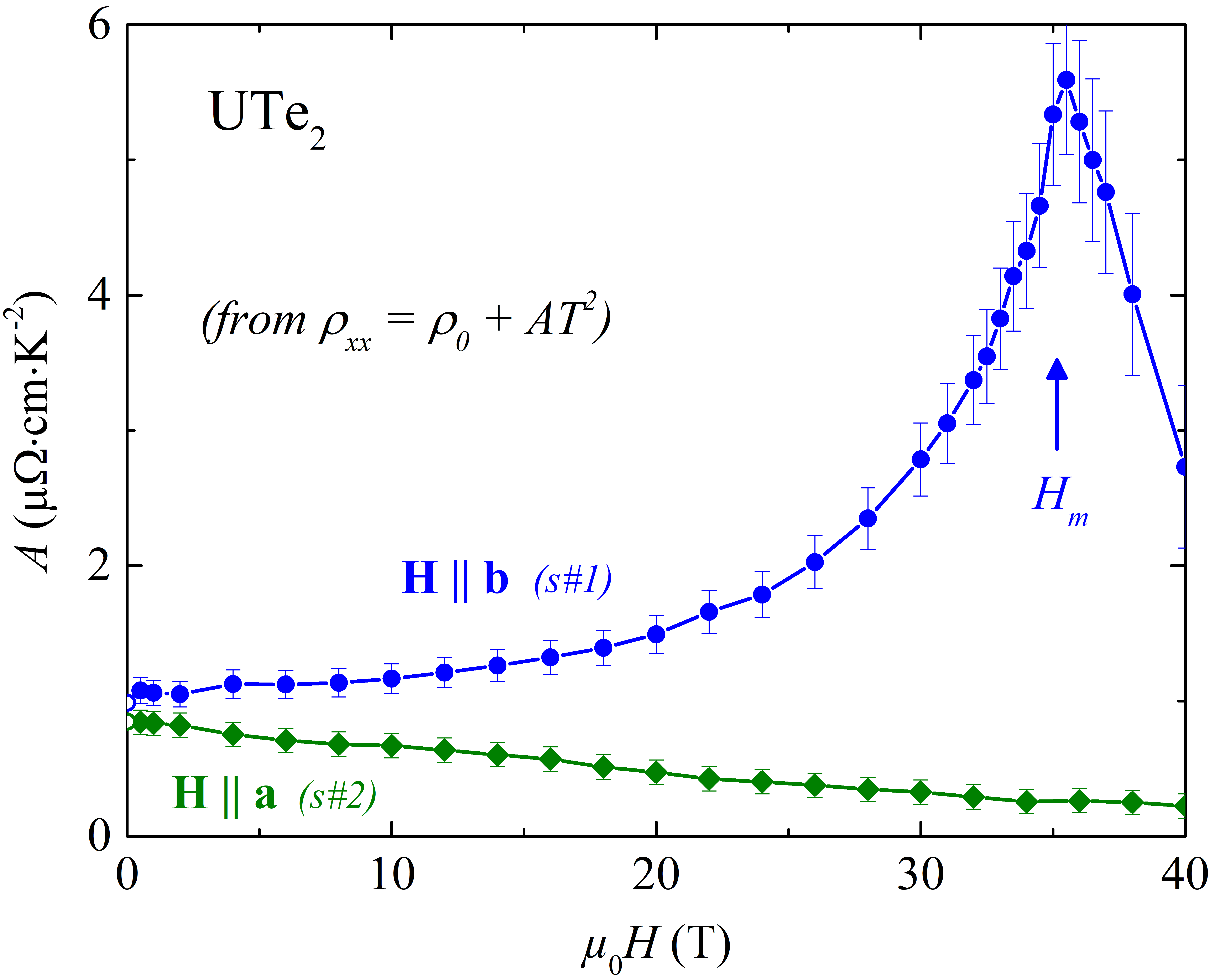}
\caption{(Color online) Field-dependence of the quadratic coefficient $A$ of the electrical resistivity for $\mathbf{H} \parallel \mathbf{a}$ and  $\mathbf{b}$.}
\label{Fig4}
\end{figure}

Figure \ref{Fig4} shows the field-dependence of the quadratic term $A$ extracted from a Fermi-liquid fit, for $T\leq4.2$~K, to the electrical resistivity by $\rho_{xx}(T)=\rho_{xx}^0+AT^2$ measured for the magnetic-field directions $\mathbf{H} \parallel \mathbf{a}$ and  $\mathbf{b}$ (see details in the Supplementary Materials \cite{SM}). At a second-order quantum phase transition we expect an enhancement of magnetic fluctuations which will show up as an increase of $A$. If a Fermi-liquid picture is valid, then $A$ varies as the square $m^{*2}$ of the effective mass. However, deviations from a Fermi-liquid behavior are often observed in the vicinity of a quantum phase transition. Here, a monotonous decrease of $A$ in a magnetic field $\mu_0\mathbf{H}\parallel \mathbf{a}$ up to 60~T indicates a progressive reduction of the magnetic fluctuations. For $\mathbf{H} \parallel \mathbf{b}$, despite the occurrence of a first-order phase transition a singularity in the variation of $A$ is detected at $H_m$. The $A$ coefficient increases significantly with field, reaching a sharp maximum at $H_m$ where it is approximately 6 times larger than at zero field. We note that this analysis is only qualitative, since an orbital contribution develops in the high-field resistivity and makes deviations from a standard $T^2$ Fermi-liquid behavior (see Supplementary Materials \cite{SM}). The enhancement of $A$ may be a signature of critical ferromagnetic fluctuations associated with a Fermi surface instability at $H_m$, as observed in the highly-documented cases of CeRu$_2$Si$_2$ \cite{Raymond1998,Flouquet2004,Sato2004} and URhGe \cite{Tokunaga2015}. As in URhGe, a boost of the ferromagnetic fluctuations along the $\mathbf{b}$ direction on approaching $H_m$ may be the driving mechanism for the unusual superconducting critical field of UTe$_2$ in $\mathbf{H} \parallel \mathbf{b}$. We note that the possibility of critical magnetic fluctuations at a first-order quantum phase transition, as observed here, was recently stressed.\cite{Aoki2019,Chandra2018}

A remarkable point is the sharp step-like shape of the anomaly at $H_m$ leading to a huge increase in fields beyond $H_m$ of the low-temperature resistivity. From LDA calculations, a Kondo semi-conducting ground state with flat bands near the Fermi energy has been predicted for UTe$_{2}$,\cite{Aoki2019b} contrasting with its experimentally-observed metallic state. This discrepancy may well arise by the failure of the model to take correlations into account, and the high-field polarized state may be closer to the predicted picture. Anyway this increase of the resistivity certainly points to some significant change to the carriers and to the Fermi surface at the transition, and a further challenge will be to determine the Fermi surface of UTe$_{2}$ in magnetic fields on both sides of $H_m$. Step-like increases or decreases of the resistivity have been observed at a metamagnetic or pseudo-metamagnetic transition in several heavy-fermion antiferromagnets and paramagnets in a field applied along their easy magnetic axis (see for instance CeRu$_{2}$Si$_{2}$ and CeRh$_{2}$Si$_{2}$ \cite{Daou2006,Knafo2010}), where dramatic changes of the magnetic fluctuations (probed directly by inelastic neutron scattering \cite{Rossat1988,Raymond1998,Flouquet2004,Sato2004} or indirectly via the $A$ coefficient of the electrical resistivity) and the Fermi surface \cite{Aoki2014,Boukahil2014,Pourret2017,Gotze2017} have been reported. When the temperature is increased in these systems, the low-temperature step-like variation of $\rho(H)$ transforms into an almost symmetric and sharp peak before being replaced by a broad maximum at even higher temperatures. In the paramagnet UCoAl, which is suspected to lie in the vicinity of a ferromagnetic instability, a critical endpoint similar to that reported here for UTe$_2$ was observed at the termination of a first-order metamagnetic transition in a field applied along its easy magnetic axis\cite{Aoki2011}. Step-like variations of $\rho$, a sharp enhancement of $A$, and a change of carrier number were also found at $H_m$ in UCoAl\cite{Aoki2011,Matsuda2000,Combier2013,Palacio2013}. This indicates that the physics of UTe$_{2}$ in its CPM regime might, thus, be comparable to that of other heavy-fermions magnets, where significant roles are played by magnetic fluctuations and Fermi surface instabilities at $H_m$.

It is very likely that the large and sharp transition at $H_m$ drives the unusual superconducting properties of UTe$_{2}$ under magnetic field. The significant increase of the $A$ coefficient implies an enhancement of the magnetic fluctuations and of the effective mass at $H_m$. This is quite similar to the URhGe case, where both the superconducting temperature and the $A$ coefficient are enhanced at $H_m$ \cite{Gourgout2016,Miyake2008,Miyake2009}. In URhGe, it was also found that when the metamagnetic transition is tuned to lower fields with uniaxial stress, then superconductivity is enhanced even at zero field \cite{Braithwaite2018}. This effect was shown to be related to the concomitant increase of the transverse magnetic susceptibility. In UTe$_{2}$, the upturn of $H_{c2}$ reported for $\mathbf{H} \parallel \mathbf{b}$ by Ran et al. \cite{Ran2019} can be been interpreted as resulting from a field-enhancement of the pairing strength and could be a consequence of the nearby metamagnetic transition at $H_m$. UTe$_{2}$ is, thus, a candidate for field-induced reentrant superconductivity in the proximity of $H_m$.

In comparison with the URhGe and UCoGe cases,\cite{Miyake2008,Aoki2011b} the relative variation of $\rho_{xx}$ and $A$ at $H_m$ are sharper and stronger in UTe$_2$, their amplitude being for instance three times larger in UTe$_2$ than in URhGe. These differences at $H_m$, but also in the initial groundstates (paramagnetism and -suspected- nearby Kondo insulating state in UTe$_2$, ferromagnetism in UCoGe and URhGe) may be related to different carrier number variations and to deep changes of the Fermi surface topology, which will need to be considered. This indicates that the next challenge is to study the interplay between the magnetic, Fermi surface and superconducting properties in UTe$_2$. We end by noting that, in parallel to our work, high-field metamagnetism in UTe$_{2}$ has also been evidenced by Ran \textit{et al.} \cite{Ran2019b} and Miyake \textit{et al.} \cite{Miyake2019}.

\begin{acknowledgment}

This work at the LNCMI was supported by the 'Programme Investissements d'Avenir' under the project ANR-11-IDEX-0002-02 (reference ANR-10-LABX-0037-NEXT). We acknowledge the financial support of the Cross-Disciplinary Program on Instrumentation and Detection of CEA, the French Alternative Energies and Atomic Energy Commission, and KAKENHI (JP15H05882, JP15H05884, JP15K21732, JP16H04006, JP15H05745, JP19H00646). We acknowledge discussions with A. Miyake at ISSP-Tokyo and M. Nardone, A. Zitouni and J. B\'{e}ard at LNCMI-Toulouse.

\end{acknowledgment}


\begin{thebibliography}{9}
\bibitem{Saxena2000} S. S. Saxena, P. Agarwal, K. Ahilan, F. M. Grosche, R. K. W. Haselwimmer, M. J. Steiner, E. Pugh, I. R. Walker, S. R. Julian, P. Monthoux, G. G. Lonzarich, A. Huxley, I. Sheikin, D. Braithwaite and J. Flouquet, Nature \textbf{406}, 587 (2000).
\bibitem{Aoki2001} D. Aoki, A. Huxley, E. Ressouche, D. Braithwaite, J. Flouquet, J.-P. Brison, E. Lhotel and C. Paulsen, Nature \textbf{413}, 613 (2001).
\bibitem{Huy2007} N. T. Huy, A. Gasparini, D. E. de Nijs, Y. Huang, J. C. P. Klaasse, T. Gortenmulder, A. de Visser, A. Hamann, T. G\"{o}rlach and H. v. L\"{o}hneysen, Phys. Rev. Lett. \textbf{99}, 067006 (2007).
\bibitem{Akazawa2004} T. Akazawa, H. Hidaka, T. Fujiwara, T. C. Kobayashi, E. Yamamoto, Y. Haga, R. Settai and Y. \={O}nuki, J. Phys.: Condens. Matter \textbf{16}, L29 (2004).
\bibitem{Aoki2012} D. Aoki and J. Flouquet, J. Phys. Soc. Jpn. \textbf{81}, 011003 (2012).
\bibitem{Aoki2019} D. Aoki, K. Ishida and J. Flouquet, J. Phys. Soc. Jpn. \textbf{88}, 022001 (2019).
\bibitem{Wu2017} B. Wu, G. Bastien, M. Taupin, C. Paulsen, L. Howald, D. Aoki, and J.-P. Brison, Nat. Commun. \textbf{8}, 14480 (2017).
\bibitem{Levy2005} F. L\'{e}vy, I. Sheikin, B. Grenier and A. D. Huxley, Science \textbf{309}, 1343 (2005).
\bibitem{Hardy2011} F. Hardy, D. Aoki, C. Meingast, P. Schweiss, P. Burger, H. v. L\"{o}hneysen and J. Flouquet, Phys. Rev. B \textbf{83}, 195107 (2011).
\bibitem{Yelland2011} E. A. Yelland, J. M. Barraclough, W. Wang, K. V. Kamenev and A. D. Huxley, Nat. Physics \textbf{7}, 890 (2011).
\bibitem{Gourgout2016} A. Gourgout, A. Pourret, G. Knebel, D. Aoki, G. Seyfarth and J. Flouquet, Phys. Rev. Lett. \textbf{117}, 046401 (2016).
\bibitem{Miyake2008} A. Miyake, D. Aoki and J. Flouquet, J. Phys. Soc. Jpn. \textbf{77}, 094709 (2008).
\bibitem{Miyake2009} A. Miyake, D. Aoki and J. Flouquet, J. Phys. Soc. Jpn. \textbf{78}, 063703 (2009)
\bibitem{Aoki2009} D. Aoki, T. D. Matsuda, V. Taufour, E. Hassinger, G. Knebel and J. Flouquet, J. Phys. Soc. Jpn. \textbf{78}, 113709 (2009).
\bibitem{Knafo2012} W. Knafo, T. D. Matsuda, D. Aoki, F. Hardy, G. W. Scheerer, G. Ballon, M. Nardone, A. Zitouni, C. Meingast and J. Flouquet, Phys. Rev. B \textbf{86}, 184416 (2012).
\bibitem{Ran2019} S. Ran, C. Eckberg, Q.-P. Ding, Y. Furukawa, T. Metz, S. R. Saha, I.-L. Liu, M. Zic, H. Kim, J. Paglione and N. P. Butch, arXiv:1811.11808.
\bibitem{Ikeda2006} S. Ikeda, H. Sakai, D. Aoki, Y. Homma, E. Yamamoto, A. Nakamura, Y. Shiokawa, Y. Haga and Y. Onuki, J. Phys. Soc. Jpn. \textbf{75}, 116 (2006).
\bibitem{Sechovsky1998} V. Sechovsky and L. Havela, in Handbook of Magnetic Materials, ed. K. H. J. Buschow (Elsevier), Vol. 11, Chap. 1, p. 1 (1998).
\bibitem{Aoki2019b} D. Aoki, A. Nakamura, F. Honda, D. Li, Y. Homma, Y. Shimizu, Y. J. Sato, G. Knebel, J.-P. Brison, A. Pourret,
D. Braithwaite, G. Lapertot, Q. Niu, M. Valiska, H. Harima, and J. Flouquet, J. Phys. Soc. Jpn. \textbf{88}, 043702 (2019).
\bibitem{Miyake2019} A. Miyake, Y. Shimizu, Y.J. Sato, D. Lin, A. Nakamura, Y. Homma, F. Honda, J. Flouquet, M. Tokunaga, D. Aoki, private communication.
\bibitem{SM}  (Supplemental material) Complementary data from our high-magnetic-field study of UTe$_2$ is provided online.
\bibitem{Aoki2013} D. Aoki, W. Knafo and I. Sheikin: C. R. Phys. \textbf{14}, 53 (2013).
\bibitem{Knafo2017} W. Knafo, R. Settai, D. Braithwaite, S. Kurahashi, D. Aoki and J. Flouquet, Phys. Rev. B \textbf{95}, 014411 (2017).
\bibitem{Aoki2012b} D. Aoki, C. Paulsen, H. Kotegawa, F. Hardy, C. Meingast, P. Haen, M. Boukahil, W. Knafo, E. Ressouche, S. Raymond and J. Flouquet, J. Phys. Soc. Jpn. 81, 034711 (2012).
\bibitem{Pospisil2017} J. Posp\'{i}\v{s}il, Y. Haga, S. Kambe, Y. Tokunaga, N. Tateiwa, D. Aoki, F. Honda, A. Nakamura, Y. Homma, E. Yamamoto and T. Yamamura, Phys. Rev. B \textbf{95}, 155138 (2017).
\bibitem{Pospisil2018} J. Posp\'{i}\v{s}il, Y. Haga, Y. Kohama, A. Miyake, S. Kambe, N. Tateiwa, M. Vali\v{s}ka, P. Proschek, J. Prokle\v{s}ka, V. Sechovsk\'{y}, M. Tokunaga, K. Kindo, A. Matsuo, and E. Yamamoto, Phys. Rev. B \textbf{98}, 014430 (2018).
\bibitem{Raymond1998} S. Raymond, L. P. Regnault, S. Kambe, J. Flouquet and P. Lejay, J. Phys.: Condens. Matter \textbf{10}, 2363 (1998).
\bibitem{Flouquet2004} J. Flouquet, Y. Haga, P. Haen, D. Braithwaite, G. Knebel, S. Raymond, and S. Kambe, J. Magn. Magn. Mater. \textbf{272-276}, 27 (2004).
\bibitem{Sato2004} M. Sato, Y. Koike, S. Katano, N. Metoki, H. Kadowaki and S. Kawarazaki, J. Phys. Soc. Jpn. \textbf{73}, 3418 (2004).
\bibitem{Tokunaga2015} Y. Tokunaga, D. Aoki, H. Mayaffre, S. Kra\"{a}mer, M.-H. Julien, C. Berthier, M. Horvatic, H. Sakai, S. Kambe and S. Araki, Phys. Rev. Lett. \textbf{114}, 216401 (2015).
\bibitem{Chandra2018} P. Chandra, P. Coleman, M. A. Continentino, and G. G. Lonzarich, arXiv:1805.11771.
\bibitem{Daou2006} R. Daou, C. Bergemann and S. R. Julian, Phys. Rev. Lett. \textbf{96}, 026401 (2006).
\bibitem{Knafo2010} W. Knafo, D. Aoki, D. Vignolles, B. Vignolle, Y. Klein, C. Jaudet, A. Villaume, C. Proust and J. Flouquet, Phys. Rev. B \textbf{81}, 094403 (2010).
\bibitem{Rossat1988} J. Rossat-Mignod, L. Regnault, J. Jacoud, C. Vettier, P. Lejay, J. Flouquet, E. Walker, D. Jaccard and A. Amato, J. Magn. Magn. Mater. \textbf{76-77}, 376 (1988).
\bibitem{Aoki2014} H. Aoki, N. Kimura and T. Terashima, J. Phys. Soc. Jpn. \textbf{83}, 072001 (2014).
\bibitem{Boukahil2014} M. Boukahil, A. Pourret, G. Knebel, D. Aoki, Y. Onuki, and J. Flouquet, Phys. Rev. B \textbf{90}, 075127 (2014).
\bibitem{Pourret2017} A. Pourret, M.-T. Suzuki, A. Palaccio Morales, G. Seyfarth, G. Knebel, D. Aoki and J. Flouquet, J. Phys. Soc. Jpn. \textbf{86}, 084702 (2017).
\bibitem{Gotze2017} K. G\"{o}tze, D. Aoki, F. L\'{e}vy-Bertrand, H. Harima and I. Sheikin, Phys. Rev. B \textbf{95} 161107 (2017).
\bibitem{Aoki2011} D. Aoki, T. Combier, V. Taufour, T.D. Matsuda, G. Knebel, H. Kotegawa, and J. Flouquet, J. Phys. Soc. Jpn. \textbf{80}, 094711 (2011).
\bibitem{Matsuda2000} T.D. Matsuda, H. Sugawara, Y. Aoki, H. Sato, A.V. Andreev, Y. Shiokawa, V. Sechovsky, and L. Havela, Phys. Rev. B \textbf{62}, 13852 (2000)
\bibitem{Combier2013} T. Combier, D. Aoki, G. Knebel, and J. Flouquet, J. Phys. Soc. Jpn. \textbf{82}, 104705 (2013).
\bibitem{Palacio2013} A.Palacio-Morales, A.Pourret, G.Knebel, T.Combier, D.Aoki, H.Harima, and J.Flouquet, Phys. Rev. Lett. \textbf{110}, 116404 (2013).
\bibitem{Braithwaite2018} D. Braithwaite, D. Aoki, J.-P. Brison, J. Flouquet, G. Knebel, A. Nakamura and A. Pourret, Phys. Rev. Lett. \textbf{120}, 037001 (2018).
\bibitem{Aoki2011b} D. Aoki, T.D. Matsuda, F. Hardy, C. Meingast, V. Taufour, E. Hassinger, I. Sheikin, C. Paulsen, G. Knebel, H. Kotegawa, and J. Flouquet, J. Phys. Soc. Jpn. \textbf{80}, SA008 (2011).
\bibitem{Ran2019b} S. Ran, private communication.



\end{thebibliography}
\end{document}

\maketitle

\thispagestyle{empty}
\pagestyle{plain}

Complementary data from our high-magnetic-field study of UTe$_2$ are presented here.

\begin{figure}[h]
\makeatletter
\renewcommand{\thefigure}{S\@arabic\c@figure}
\makeatother
\begin{center}
\includegraphics[width=.6\columnwidth]{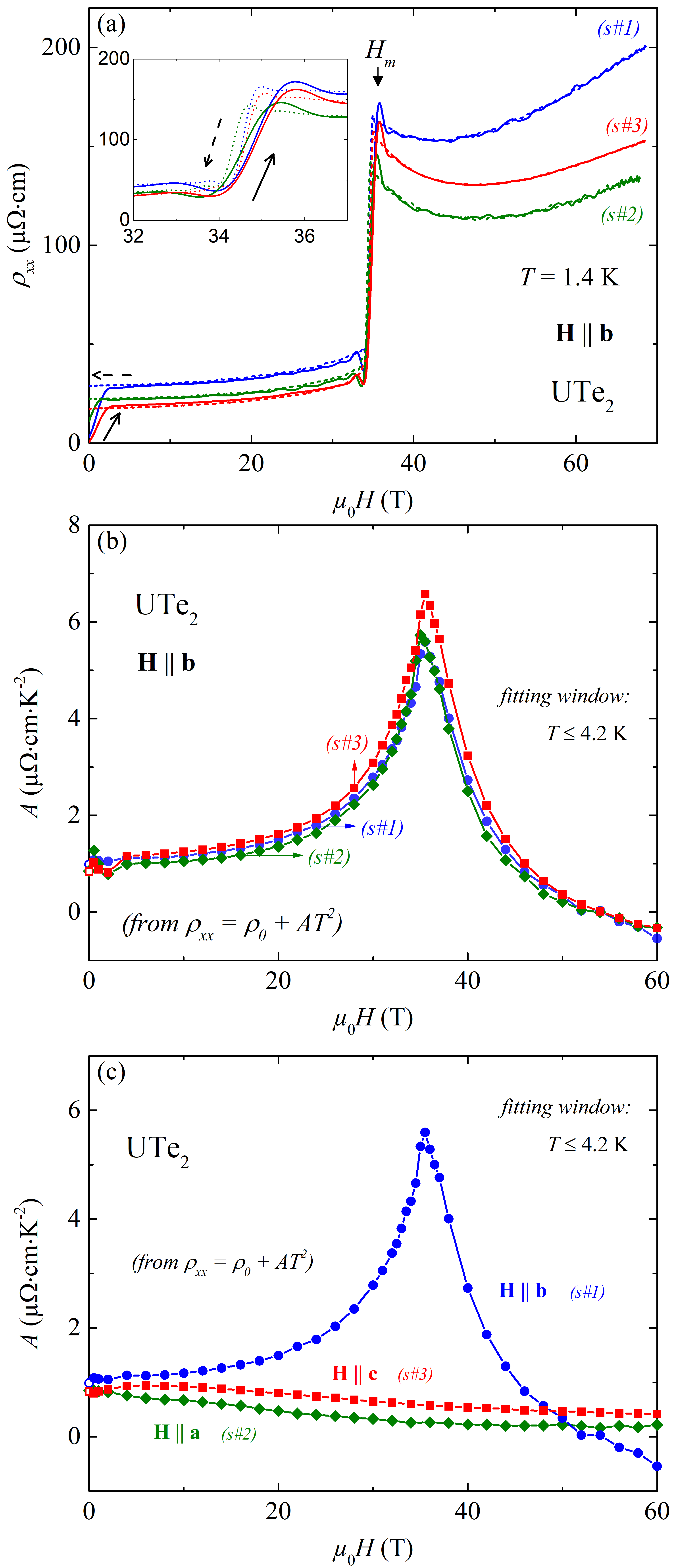}
\end{center}
\caption{(a) Resistivity versus field of UTe$_{2}$ samples $\sharp1$, $\sharp2$, and $\sharp3$ at $T = 1.4$~K in a magnetic field $\mu_0\mathbf{H} \parallel \mathbf{b}$ up to 68~T. Inset focuses on fields close to $H_m$. Field-up (solid lines) and field-down (dotted line) sweeps are plotted. (b) Field-dependence of $A$ of the electrical resistivity of samples $\sharp1$, $\sharp2$, and $\sharp3$ in a magnetic field $\mathbf{H} \parallel \mathbf{b}$ (field-up data). (c) Field-dependence of $A$ for $\mathbf{H} \parallel \mathbf{a}$ (sample $\sharp2$), $\mathbf{H} \parallel \mathbf{b}$ (sample $\sharp1$), and $\mathbf{H} \parallel \mathbf{c}$ (sample $\sharp3$) (field-up data).}
\label{FigS1}
\end{figure}

Figure \ref{FigS1}(a) compares the low-temperature ($T=1.4$~K) resistivity $\rho_{xx}$ versus magnetic field of the three samples $\sharp1$, $\sharp2$, and  $\sharp3$ (from the same batch) in $\mathbf{H} \parallel \mathbf{b}$. The three sets of data show similar features: i) a metamagnetic transition at $\mu_0H_m\simeq35$~T, ii) a similar hysteresis, of width $\Delta H \simeq 0.5$~T, at the first order transition at $H_m$, iii) a high-field orbital enhancement of $\rho_{xx}$ controlled the field-induced cyclotron motion of carrier, iv) a signature of self-heating by eddy currents at the end of the pulsed-field shot, where the sample remains in its normal non superconducting state. Figure \ref{FigS1}(b) shows the field-variation of the Fermi liquid quadratic coefficient $A$ extracted from a $\rho_{xx}(T)=\rho_{xx}^0+AT^2$ fit for the three samples in $\mathbf{H} \parallel \mathbf{b}$. The fits were made in the temperature windows $2.2\leq T\leq4.2$~K for $\mu_0H<5$~T and $T\leq4.2$~K for $\mu_0H>5$~T. Very similar variations of $A$ are obtained. Non-physical negative values of $A$ are obtained at high-field, where the orbital contribution to $\rho_{xx}$ develops and drives to a deviation from a Fermi-liquid like resistivity. Figure \ref{FigS1}(c) compares the field-variation of $A$ extracted for $\mathbf{H} \parallel \mathbf{a}$ (sample $\sharp2$), $\mathbf{H} \parallel \mathbf{b}$ (sample $\sharp1$), and $\mathbf{H} \parallel \mathbf{c}$ (sample $\sharp3$). The field-induced decrease of $A$ for $\mathbf{H} \parallel \mathbf{c}$, as well as that of $A$ for $\mathbf{H} \parallel \mathbf{b}$ in fields well above $H_m$ is presumably mainly controlled by the orbital contribution to the resistivity. These features are considered with more details in the following.

\begin{figure}[h]
\makeatletter
\renewcommand{\thefigure}{S\@arabic\c@figure}
\makeatother
\begin{center}
\includegraphics[width=.6\columnwidth]{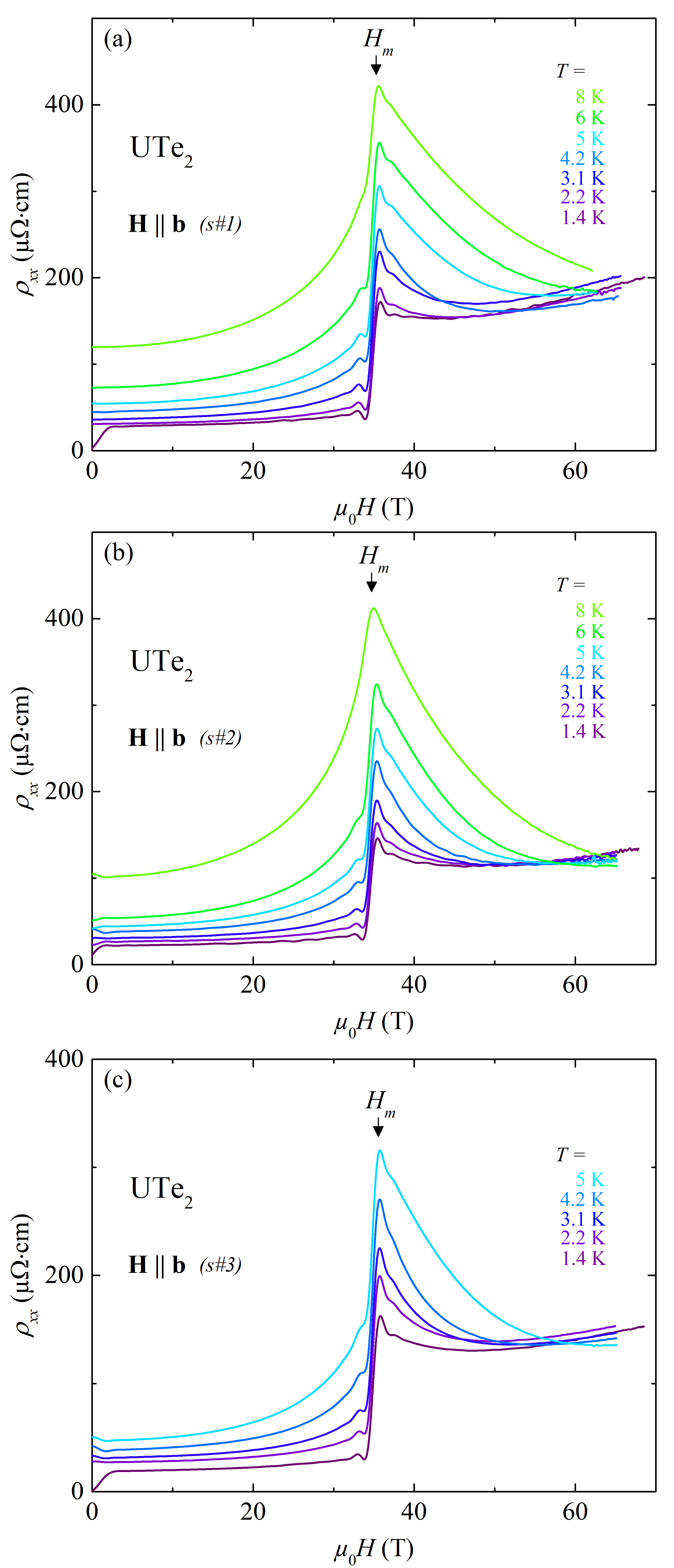}
\end{center}
\caption{Resistivity versus field of UTe$_{2}$ samples (a) $\sharp1$, (b) $\sharp2$, and (c) $\sharp3$ in a magnetic field $\mu_0\mathbf{H} \parallel \mathbf{b}$ up to 68~T, at temperatures from 1.4 to 8~K. Only field-up sweeps are shown.}
\label{FigS2}
\end{figure}

Figure \ref{FigS2} compares the magnetoresistivity of the three samples measured under temperatures from 1.4 to 8~K in a magnetic field $\mathbf{H} \parallel \mathbf{b}$. Small differences between the three sets of data are suspected to result from slightly-different sample qualities and possible small mis-orientations. Due to the use of high-frequencies $f\simeq20-70$~kHz in our four-point resistivity measurements under pulsed magnetic fields, small out-of-phase contaminations due to parasitic capacitances (for instance at the electrical contacts on the surface of the samples) also affect the absolute variations of $\rho_{xx}$ and limit the reproducibility of the measurements.

\begin{figure}[h]
\makeatletter
\renewcommand{\thefigure}{S\@arabic\c@figure}
\makeatother
\begin{center}
\includegraphics[width=.6\columnwidth]{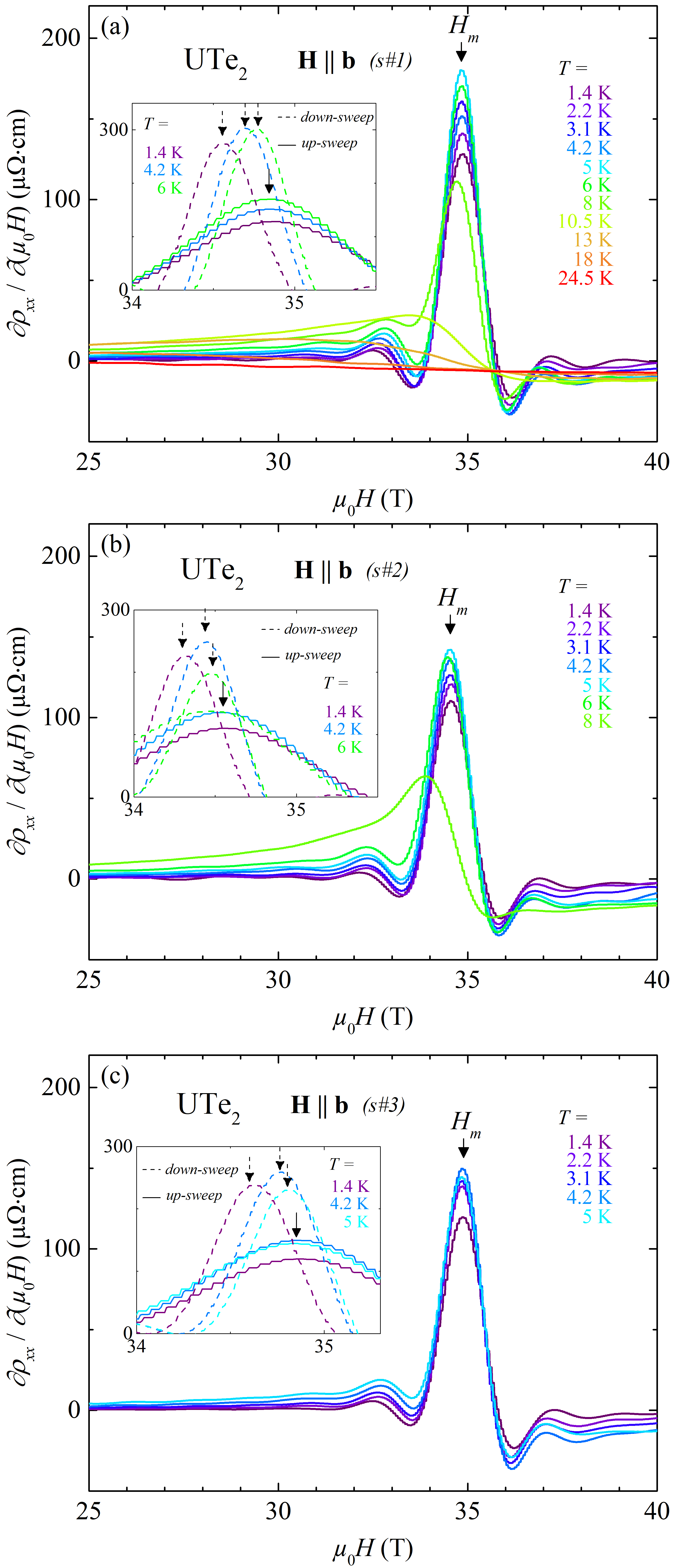}
\end{center}
\caption{Field-derivative of resistivity versus field of UTe$_{2}$ samples (a) $\sharp1$, (b) $\sharp2$, and (c) $\sharp3$ in a magnetic field $\mu_0\mathbf{H} \parallel \mathbf{b}$ up to 68~T, at temperatures from 1.4 to 24.5~K. Only field-up sweeps are shown in the main panels of the graphs. Zooms on the comparison between field-up and field-down sweeps in the vicinity of $Hm$ are shown in the Insets.}
\label{FigS3}
\end{figure}

Figure \ref{FigS3} shows the field-derivative of the resistivity of the three samples at different temperatures in $\mathbf{H} \parallel \mathbf{b}$. Variations of the metamagnetic field $\mu_0H_m=34.55-34.9$~T (field-up sweeps), estimated at the maximum of $\partial\rho_{xx}/\partial H$, probably result from small mis-orientations of the samples. Oscillations on both sides of the main peak at $H_m$ are suspected to be non-physical artefacts due to the digital lock-in treatment and the sharpness and amplitude of the anomaly. Insets in Figure \ref{FigS3} compare field-up and field-down sweeps, in the vicinity of $H_m$, and show the hysteresis characterizing this first-order transition present at temperatures $T<T_{CEP}=7$~K. Similar hysteresis are obtained for the three samples studied here.

\begin{figure}[h]
\makeatletter
\renewcommand{\thefigure}{S\@arabic\c@figure}
\makeatother
\begin{center}
\includegraphics[width=.6\columnwidth]{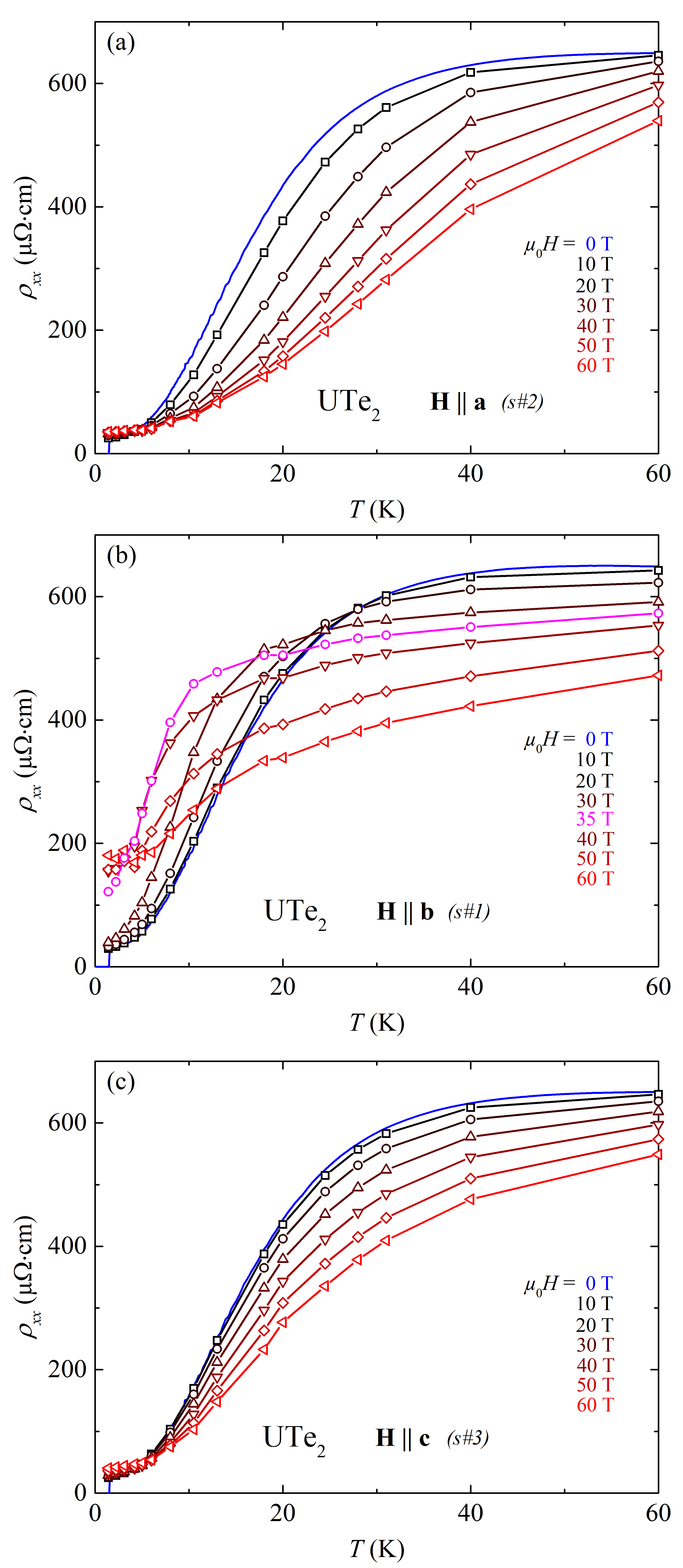}
\end{center}
\caption{Resistivity versus temperature of UTe$_{2}$ in magnetic fields (a) $\mu_0\mathbf{H} \parallel \mathbf{a}$ (sample $\sharp2$),  $\mu_0\mathbf{H} \parallel \mathbf{b}$ (sample $\sharp1$), and $\mu_0\mathbf{H} \parallel \mathbf{c}$ (sample $\sharp3$) up to 60~T. Only field-up sweeps are shown.}
\label{FigS4}
\end{figure}

Figure \ref{FigS4} shows how a magnetic field leads to a progressive disappearance of the high-temperature broad maximum in the electrical resistivity and, thus, of the associated electronic correlations signatures of the correlated paramagnetic (CPM) regime. For $\mathbf{H} \parallel \mathbf{a},\mathbf{c}$, a small field-induced enhancement of $\rho_{xx}$ is visible at low temperature, and $\rho_{xx}$ monotonously decreases with field at temperatures $T\geq8$~K. For $\mathbf{H} \parallel \mathbf{b}$, additional features are observed at low temperature, in relation with the sharp increase of $\rho_{xx}$ at $H_m$, while a high-temperature monotonous decrease with field - similar to those observed for the other field directions - is observed at temperatures $T\geq25$~K. For $\mathbf{H} \parallel \mathbf{b}$, a scattering of the $\rho_{xx}$ versus $T$ data (extracted at different fields from our pulsed field experiments) at low temperature and high-field indicates the limits of reproducibility of the measurements.

\begin{figure}[h]
\makeatletter
\renewcommand{\thefigure}{S\@arabic\c@figure}
\makeatother
\begin{center}
\includegraphics[width=.6\columnwidth]{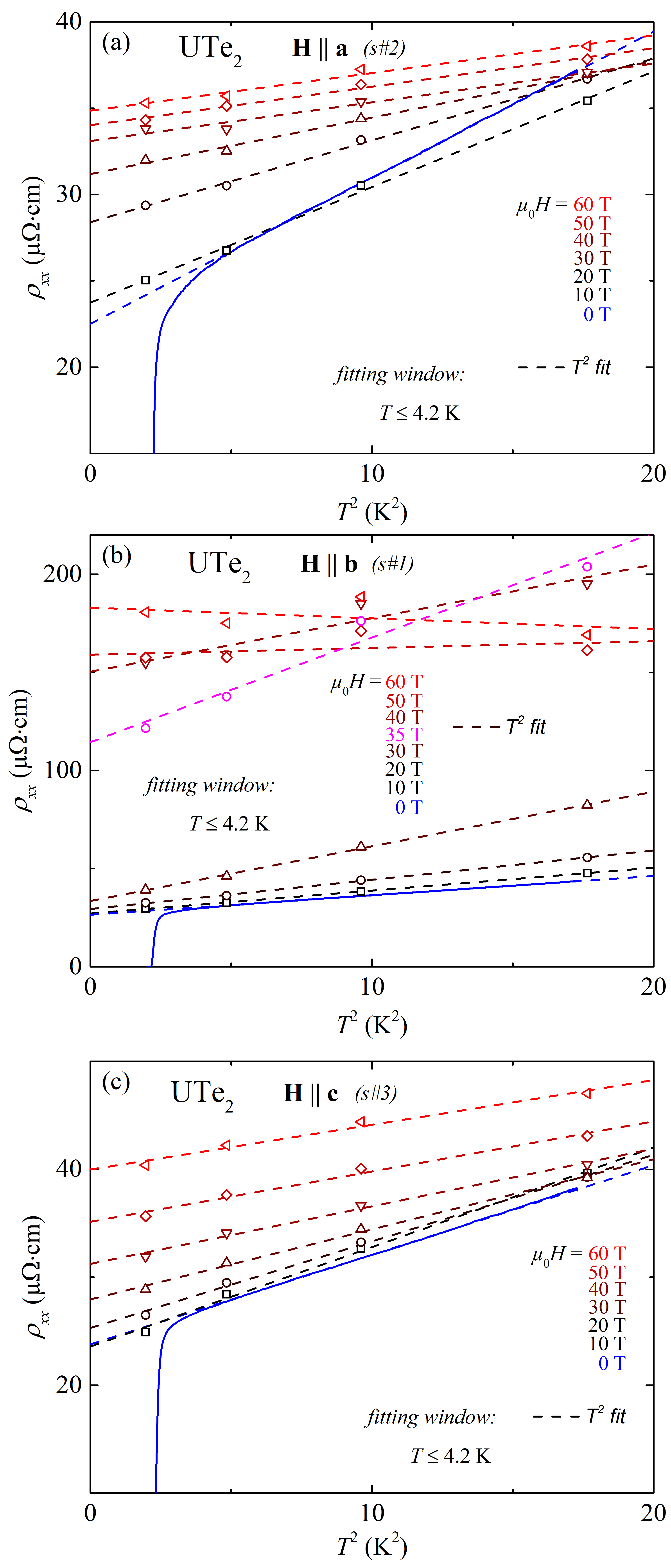}
\end{center}
\caption{Resistivity versus square of temperature of UTe$_{2}$ and its fit by $\rho_{xx}(T)=\rho_{xx}^0+AT^2$, extracted for $T\leq4.2$~K, (dotted lines) in magnetic fields (a) $\mu_0\mathbf{H} \parallel \mathbf{a}$ (sample $\sharp2$),  $\mu_0\mathbf{H} \parallel \mathbf{b}$ (sample $\sharp1$), and $\mu_0\mathbf{H} \parallel \mathbf{c}$ (sample $\sharp3$) up to 60~T. Only field-up sweeps are shown.}
\label{FigS5}
\end{figure}

\begin{figure}[h]
\makeatletter
\renewcommand{\thefigure}{S\@arabic\c@figure}
\makeatother
\begin{center}
\includegraphics[width=.6\columnwidth]{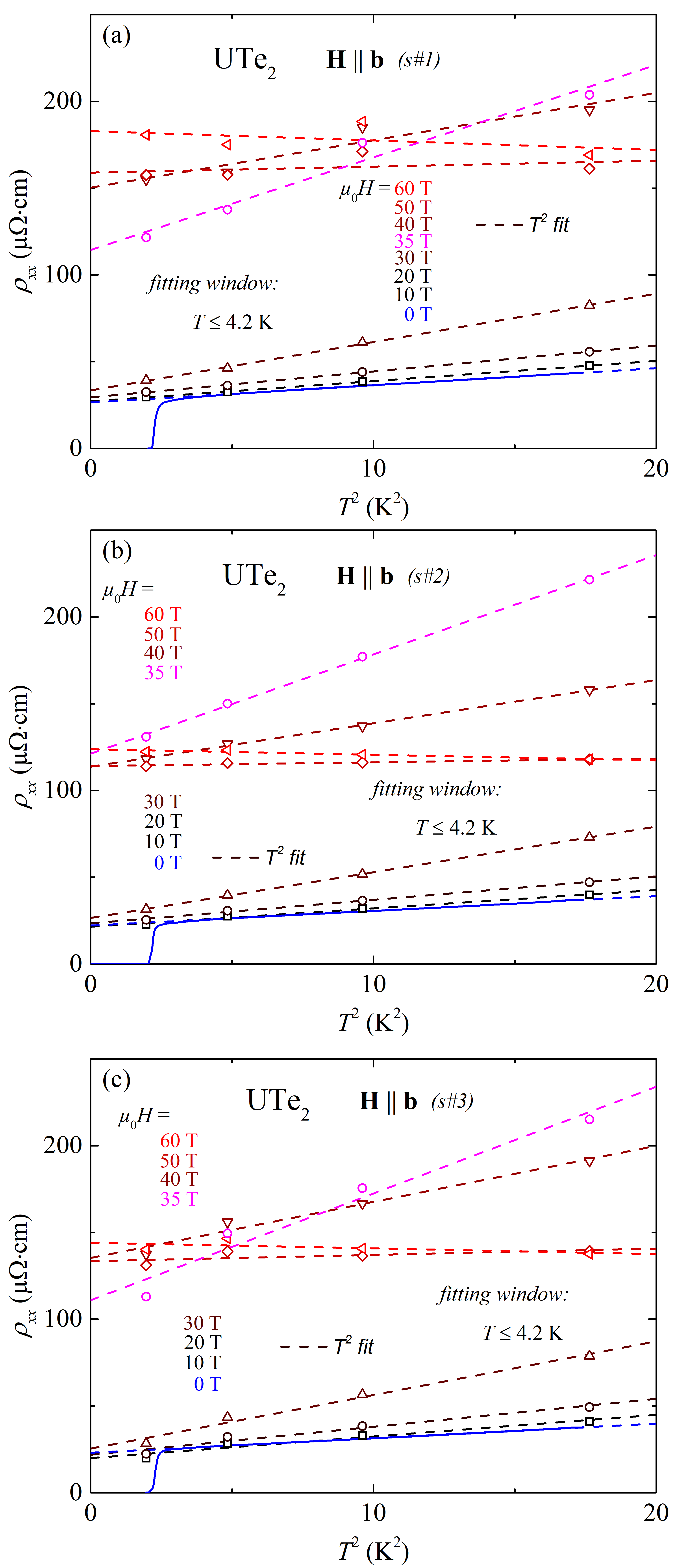}
\end{center}
\caption{Resistivity versus square of temperature of UTe$_{2}$ and its fit by $\rho_{xx}(T)=\rho_{xx}^0+AT^2$, extracted for $T\leq4.2$~K, (dotted lines) in magnetic fields $\mu_0\mathbf{H} \parallel \mathbf{b}$ up to 60~T for (a) sample $\sharp1$,  (b) sample $\sharp2$), and (c) sample $\sharp3$. Only field-up sweeps are shown.}
\label{FigS6}
\end{figure}

Figure \ref{FigS5} shows details of the fit to the resistivity data by $\rho_{xx}(T)=\rho_{xx}^0+AT^2$ for $T\leq4.2$~K, at several field values from 0 to 60~T, with $\mathbf{H} \parallel \mathbf{a}$ (sample $\sharp1$), $\mathbf{b}$ (sample $\sharp2$), and  $\mathbf{c}$ (sample $\sharp3$). Figure \ref{FigS6} shows details of the fit to the resistivity data by $\rho_{xx}(T)=\rho_{xx}^0+AT^2$ for $T\leq4.2$~K, at several field values from 0 to 60~T, for samples $\sharp1$, $\sharp2$, and  $\sharp3$ with $\mathbf{H} \parallel \mathbf{b}$.

\begin{figure}[h]
\makeatletter
\renewcommand{\thefigure}{S\@arabic\c@figure}
\makeatother
\begin{center}
\includegraphics[width=.6\columnwidth]{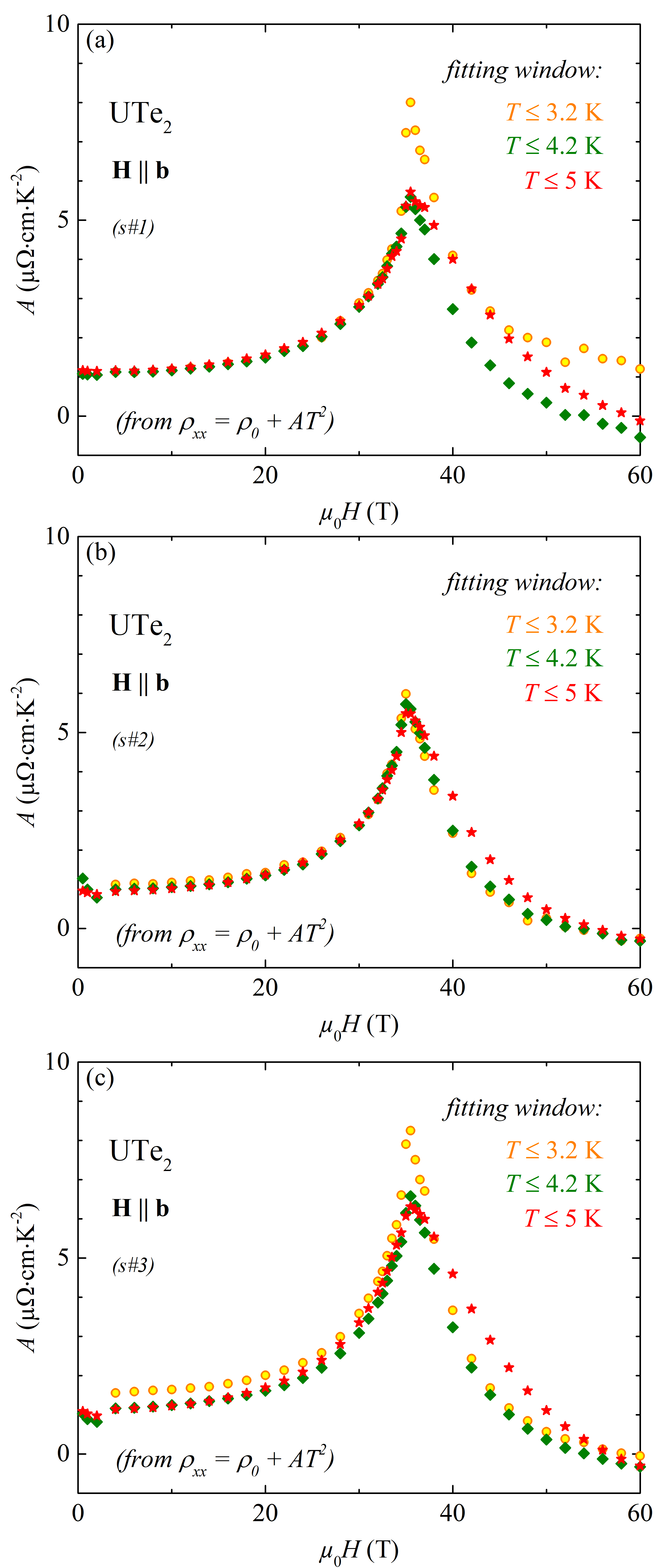}
\end{center}
\caption{Field-dependence of the quadratic coefficient $A$, extracted for temperatures windows $T\leq3.2$, 4.2, and 5~K, of the electrical resistivity of UTe$_{2}$ in magnetic fields $\mu_0\mathbf{H} \parallel \mathbf{b}$ up to 60~T for (a) sample $\sharp1$,  (b) sample $\sharp2$), and (c) sample $\sharp3$. Only field-up sweeps are shown.}
\label{FigS7}
\end{figure}

Figure \ref{FigS7} presents the field-dependence of the quadratic coefficient $A$ extracted for different temperatures windows $T\leq3.2$, 4.2, and 5~K, for the three samples $\sharp1$, $\sharp2$, and  $\sharp3$ with $\mathbf{H} \parallel \mathbf{b}$. It shows that $A$ is quite sensitive to the choice of the field-window. This is due to the scattering of our $\rho_{xx}$ versus $T$ data visible in Figures \ref{FigS2}, \ref{FigS4}, \ref{FigS5}, and  \ref{FigS6}. We find that our data, for the three field-directions and the three samples, are compatible with a $T^2$ variation of $\rho_{xx}$ in the temperature windows $2.2\leq T\leq4.2$~K for $\mu_0H<5$~T (due to the low-temperature superconducting phase) and $T\leq4.2$~K for $\mu_0H>5$~T. These windows were used to extract the $A$ coefficient variations shown in Figure \ref{FigS1}(b) and Figure 4 of the paper.

\begin{figure}[h]
\makeatletter
\renewcommand{\thefigure}{S\@arabic\c@figure}
\makeatother
\begin{center}
\includegraphics[width=.6\columnwidth]{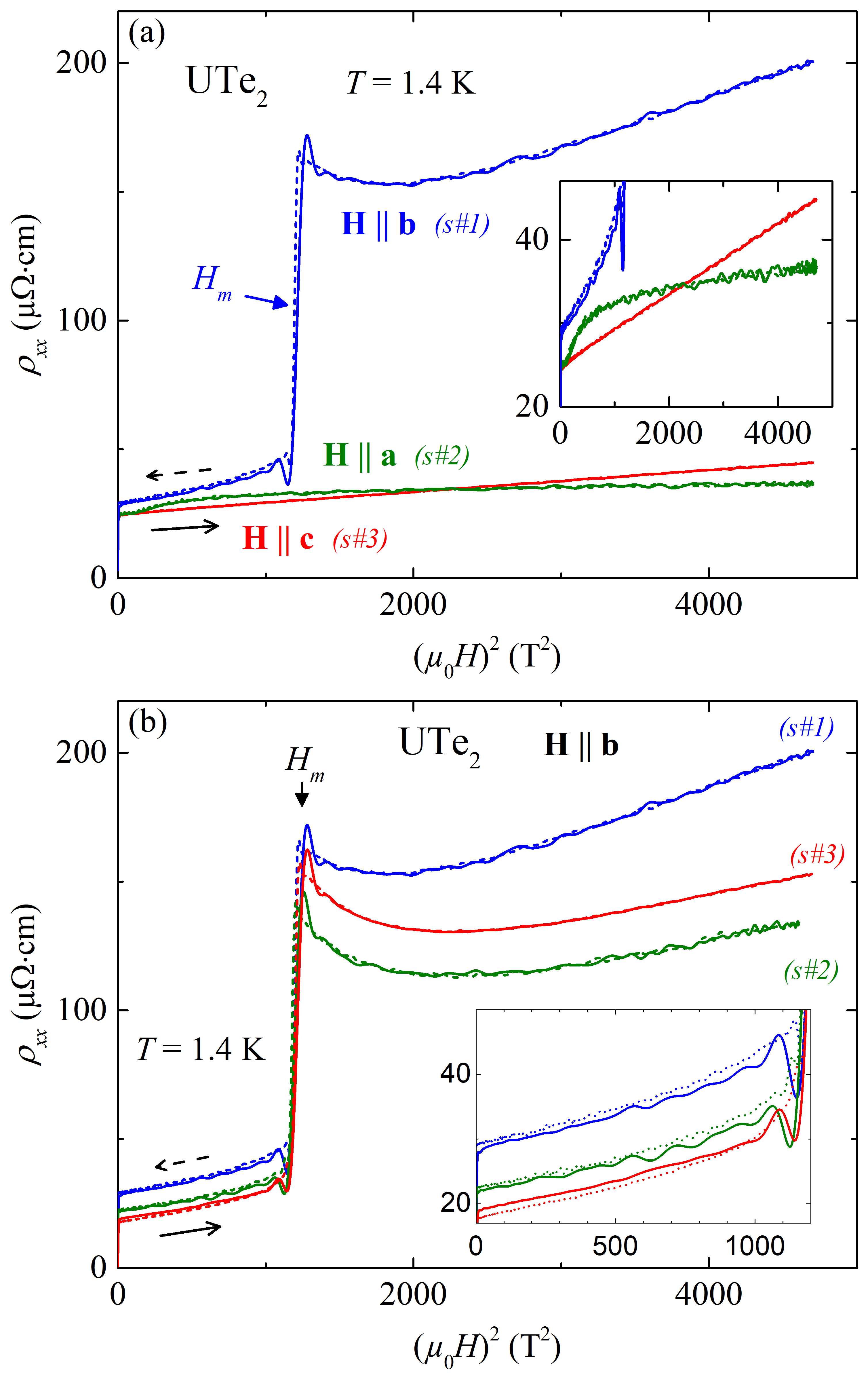}
\end{center}
\caption{(a) Resistivity versus square of magnetic field of UTe$_{2}$ at $T = 1.4$~K for fields $\mu_0\mathbf{H} \parallel \mathbf{a}$ (sample $\sharp2$),  $\mu_0\mathbf{H} \parallel \mathbf{b}$ (sample $\sharp1$), and $\mu_0\mathbf{H} \parallel \mathbf{c}$ (sample $\sharp3$) up to 68~T. Inset focuses on the small resistivity variations observed for $\mathbf{H} \parallel \mathbf{a},\mathbf{c}$. (b) Resistivity versus square of magnetic field of UTe$_{2}$ at $T = 1.4$~K of samples $\sharp1$, $\sharp2$, and $\sharp3$ in a magnetic field $\mu_0\mathbf{H} \parallel \mathbf{b}$ up to 68~T. Inset focuses on the low fields $H<H_m$. Both field-up (solid lines) and field-down (dotted line) sweeps are plotted. Black arrows indicate the time direction during the field pulses.}
\label{FigS8}
\end{figure}

Figure \ref{FigS8} shows plots of $\rho_{xx}$ versus $H^2$ for the three samples and the three field directions at $T=1.4$~K. It focuses on the orbital contribution to $\rho_{xx}$. This contribution controlled by the field-induced cyclotron motion of the carriers usually leads to a $H^2$ increase of the transverse resistivity in high fields and low temperatures. For $\mathbf{H} \parallel \mathbf{c}$, a $H^2$ variation is found in the whole field window, from 0 to 68~T, and indicates that the Fermi surface is not modified. For $\mathbf{H} \parallel \mathbf{b}$, the metamagnetic transition at $H_m$ induces a deviation from a $H^2$ variation, which seems to be recovered in fields well above $H_m$. For the longitudinal configuration $\mathbf{H} \parallel \mathbf{a}$, a $H^2$ orbital variation is not expected, and a small $H^2$ contribution found in fields higher than 40~T might result from a small sample mis-orientation. The orbital contribution to $\rho_{xx}$ induces a deviation at high field from the $T^2$-dependence of the resistivity. Here, it leads to non-physical negatives values of the $A$ coefficient at fields well above $H_m$ for $\mathbf{H} \parallel \mathbf{b}$ (see Figures \ref{FigS1} and \ref{FigS7}). For this reason, for only values of $A$ under magnetic fields $\mu_0\mathbf{H} \parallel \mathbf{b}$ up to 40~T, i.e., a few T above $\mu_0H_m$, have been plotted in Figure 4 of the paper. For $\mathbf{H} \parallel \mathbf{c}$, the field-induced decrease of the $A$ coefficient (see Figure \ref{FigS8}(c)) is driven by the orbital effect and cannot be considered as the signature of decreasing effective mass and magnetic fluctuations (see Figure \ref{FigS5}). For this reason, the values of $A$ under magnetic fields $\mathbf{H} \parallel \mathbf{c}$ have not been plotted in Figure 4 of the paper.

\begin{figure}[h]
\makeatletter
\renewcommand{\thefigure}{S\@arabic\c@figure}
\makeatother
\begin{center}
\includegraphics[width=.6\columnwidth]{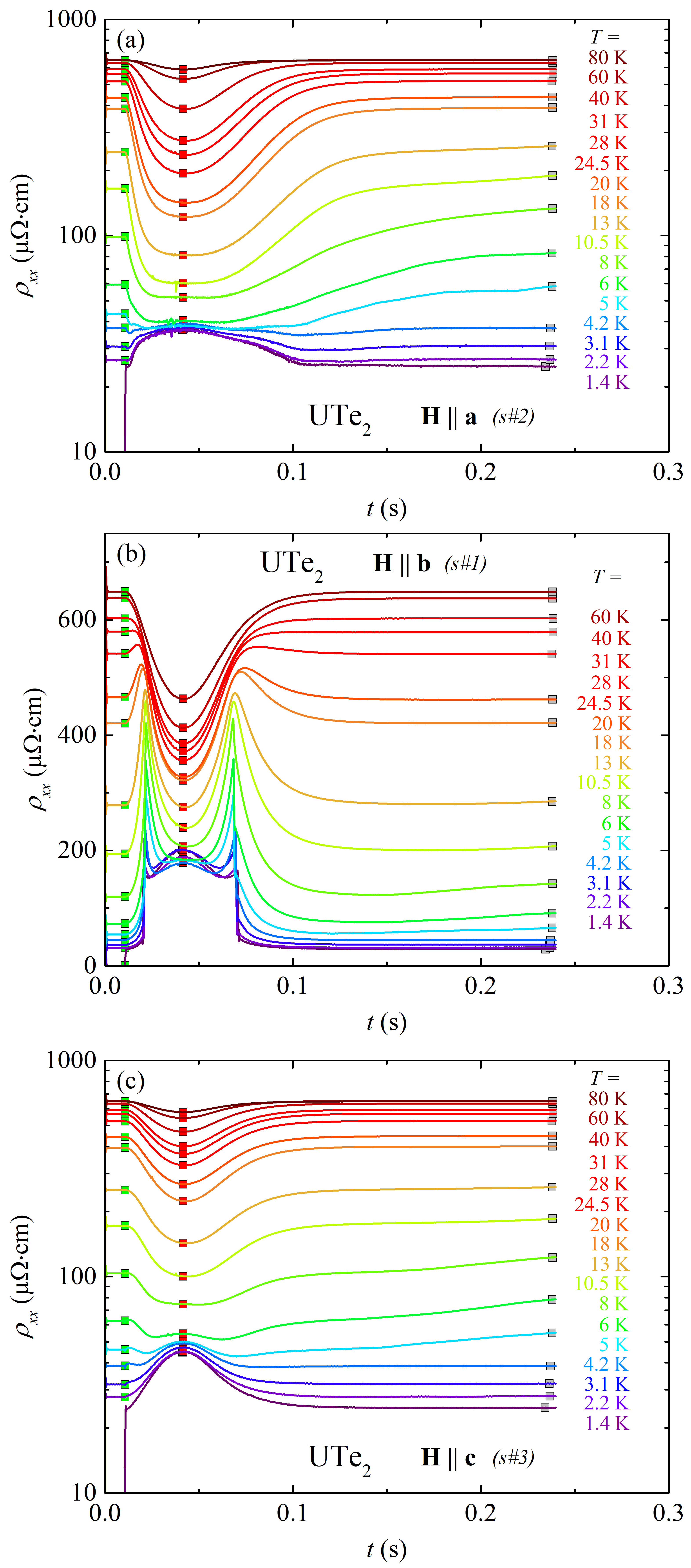}
\end{center}
\caption{Resistivity of UTe$_{2}$ versus time during magnetic fields (a) $\mu_0\mathbf{H} \parallel \mathbf{a}$ (sample $\sharp2$),  $\mu_0\mathbf{H} \parallel \mathbf{b}$ (sample $\sharp1$), and $\mu_0\mathbf{H} \parallel \mathbf{c}$ (sample $\sharp3$) pulsed up to $60-68$~T, at temperatures 1.4~$\leq T\leq$~80~K. The green, red and grey squares indicate the beginning, maximum field and end of the magnetic-field pulses, respectively.}
\label{FigS9}
\end{figure}

\begin{figure}[h]
\makeatletter
\renewcommand{\thefigure}{S\@arabic\c@figure}
\makeatother
\begin{center}
\includegraphics[width=.57\columnwidth]{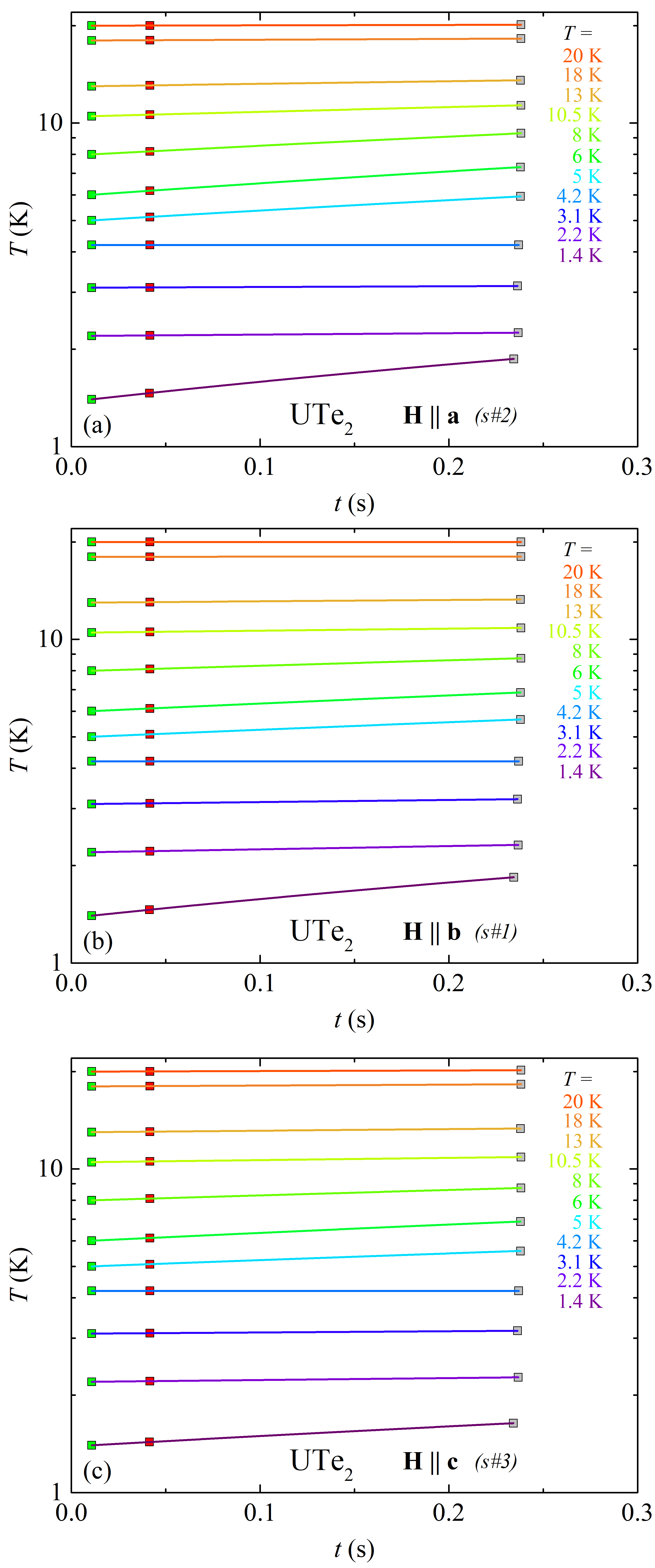}
\end{center}
\caption{Temperature of UTe$_{2}$ estimated during the magnetic-field pulses for (a) $\mathbf{H} \parallel \mathbf{a}$ (sample $\sharp2$),  $\mathbf{H} \parallel \mathbf{b}$ (sample $\sharp1$), and $\mathbf{H} \parallel \mathbf{c}$ (sample $\sharp3$), at temperatures 1.4~$\leq T\leq$~20~K. The increase of temperature at the end of the pulsed has been estimated from the resistance variation before and after the pulse, and a linear variation of temperature versus time has been assumed. The green, red and grey squares indicate the beginning, maximum field and end of the magnetic-field pulse, respectively.}
\label{FigS10}
\end{figure}

Figure \ref{FigS9} shows the time dependence of the resistivity extracted during the different pulsed-fields shots done in this work, with magnetic fields $\mu_0\mathbf{H} \parallel \mathbf{a}$ (sample $\sharp2$),  $\mu_0\mathbf{H} \parallel \mathbf{b}$ (sample $\sharp1$), and $\mu_0\mathbf{H} \parallel \mathbf{c}$ (sample $\sharp3$) up to $60-68$~T. An increase of the zero-field resistivity at the end of the pulse, in comparison with its initial value before the pulse, is a consequence of a self-heating of the samples by eddy currents. By comparing the resistivity before and after the pulses with the temperature dependence of the resistivity measured at zero-field, the temperature of the sample at the end of the pulse is extracted (see Figure \ref{FigS10}). Assuming a linear increase of the temperature during a pulsed field shot (which is a rough approximation, but constitutes a good indication of the order of magnitude of heating effects during the pulses), we estimate the temperature of the samples at the maximum of the field.

\begin{figure}[h]
\makeatletter
\renewcommand{\thefigure}{S\@arabic\c@figure}
\makeatother
\begin{center}
\includegraphics[width=.6\columnwidth]{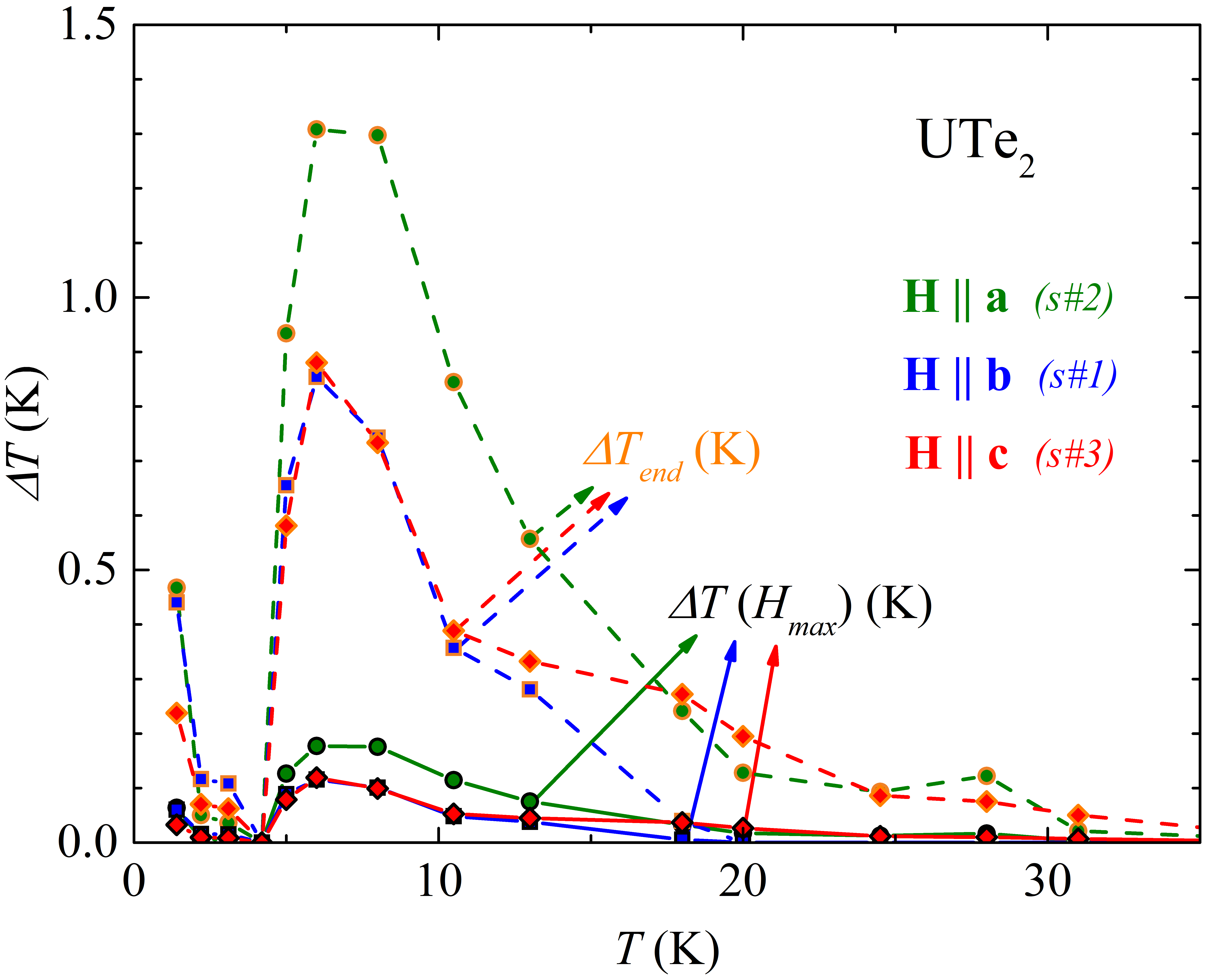}
\end{center}
\caption{Estimation of the variation of temperature $\Delta T(H_{max})$ at the maximum of the field and $\Delta T_{end}$ at the end of the pulse, for $\mathbf{H} \parallel \mathbf{a}$ (sample $\sharp2$), $\mu_0\mathbf{H} \parallel \mathbf{b}$ (sample $\sharp1$), and $\mu_0\mathbf{H} \parallel \mathbf{c}$ (sample $\sharp3$) pulsed up to 60-68~T, at temperatures 1.5~$\leq T\leq$ ~35~K.}
\label{FigS11}
\end{figure}

Figure \ref{FigS11} presents the temperature increase $\Delta T(H_{max})$ at the maximum of the field and $\Delta T_{end}$ at the end of the pulse, estimated here for the three sets of measurements with magnetic fields $\mathbf{H} \parallel \mathbf{a}$ (sample $\sharp2$), $\mu_0\mathbf{H} \parallel \mathbf{b}$ (sample $\sharp1$), and $\mu_0\mathbf{H} \parallel \mathbf{c}$ (sample $\sharp3$) pulsed up to 60-68~T, at temperatures 1.5~$\leq T\leq$ ~35~K. The variations of temperature estimated here, and summarized in Figure \ref{FigS11}, are found to be acceptable, indicating a negligible heating during up-sweeps of the magnetic field. Heating is small ($\Delta T(H_{max})<0.1$~K) for $T\leq4.2$~K, where the $A$ coefficient is extracted from $T^2$ fit to $\rho_{xx}$ from upsweep data, and very limited ($\Delta T(H_{max})<0.2$~K) at all other temperatures, where the phase diagram is extracted. More precisely, heating effects are negligible at $T=4.2$~K, where the samples are in liquid helium and where the cooling power is maximal, and for $T>30$~K, where the eddy currents are small due to bigger sample electrical resistances. Heating also remains small at temperatures $T<4.2$~K, where the samples are in depressurized liquid helium, for which the cooling power is slightly reduced in comparison with liquid helium at $T=4.2$~K (atmospheric pressure). The highest variations of temperature are found for $5\leq T\leq15$~K, where the samples are in gaseous helium, for which the cooling power is reduced in comparison with that of liquid helium, and where the sample electrical resistances are small and lead to eddy currents.

